\documentclass{ws-p8-50x6-00}
\usepackage{amsmath,epsfig}

\makeatletter
\newbox\slashbox \setbox\slashbox=\hbox{\large$/$}
\def\pslash#1{\setbox\@tempboxa=\hbox{$#1$}
  \@tempdima=0.5\wd\slashbox \advance\@tempdima 0.5\wd\@tempboxa
  \copy\slashbox \kern-\@tempdima \box\@tempboxa}
\def\slash{\protect\pslash}
\makeatother

\renewcommand\today{May 9, 1999}

\begin{document}

\title{Predictions of chiral random matrix theory and lattice QCD
  results}

\author{T. Wettig}

\address{Institut f\"ur Theoretische Physik, Technische
  Universit\"at M\"unchen,\\ D-85747 Garching, Germany}

\maketitle

\abstracts{Chiral random matrix theory makes very detailed predictions
  for the spectral correlations of the QCD Dirac operator, both in the
  bulk of the spectrum and near zero virtuality.  These predictions
  have been successfully tested in lattice QCD simulations by several
  groups.  Moreover, the domain of validity of random matrix theory
  has been predicted theoretically and identified in lattice data.
  In this talk, the current numerical evidence is reviewed.}

\section{Introduction}

We are interested in learning as much as we can about the spectrum of
the Euclidean QCD Dirac operator,
$\slash{D}=\slash{\partial}+ig\slash{A}$.  One of the reasons for this
interest is the Banks-Casher relation,
$\Sigma\equiv|\langle\bar\psi\psi\rangle|=\pi\rho(0)/V$, which relates
the chiral condensate to the spectral density,
$\rho(\lambda)=\langle\sum_n\delta(\lambda-\lambda_n)\rangle$, of the
Dirac operator at zero virtuality.  Thus, the spontaneous breaking of
chiral symmetry, a nonperturbative phenomenon with profound
consequences for the hadron spectrum, is encoded in an accumulation of
the small Dirac eigenvalues.

It has been realized that chiral random matrix theory (RMT) is a
suitable tool to compute the distribution and the correlations of the
small Dirac eigenvalues, see the talk by Verbaarschot in these
proceedings and references therein.  The first numerical evidence in
support of this statement came from instanton liquid
simulations.\cite{Verb94b} The purpose of this talk is to verify the
RMT-predictions by comparing them to lattice data.  We shall see that
the agreement between theory and numerical experiment is remarkable.

Recently, it has been shown\cite{Osbo98b} that the results obtained in
RMT can also be derived directly from field theory (partially quenched
chiral perturbation theory).  Conceptually, this is a big step
forward, while for actual calculations RMT seems to be the simpler
alternative.  It is comforting to know that the results of the two
approaches agree with each other and with lattice data.  Related
topics are covered in the talks by Damgaard, Akemann, Papp, Markum,
Stephanov, and Halasz (in order of appearance) in these proceedings.

\section{Symmetries of the Dirac operator}
\vspace*{-1mm}

Because of $\{i\slash{D},\gamma_5\}=0$, the Dirac operator falls into
one of three symmetry classes corresponding to the three chiral
ensembles of RMT: chiral orthogonal (chOE), unitary (chUE), and
symplectic (chSE) ensemble, respectively.\cite{Verb94a} In most cases,
the Dirac operator does not have additional symmetries and is
described by the chUE.  Exceptions are as follows ($N_c$ = number of
colors).

\noindent -- continuum, $N_c=2$, fermions in fundamental
representation: chOE\newline  
-- continuum, fermions in adjoint representation: chSE\newline 
-- lattice, $N_c=2$, staggered fermions in fundamental
representation: chSE\newline 
-- lattice, staggered fermions in adjoint representation: chOE

\noindent The overlap Dirac operator on the lattice has the symmetries
of the continuum operator.  There are a few exceptions where the Dirac
operator does not have the usual chiral symmetries and is, therefore,
described by the non-chiral RMT ensembles: QCD in three dimensions
(UE), and the Wilson Dirac operator on the lattice (OE for $N_c=2$, UE
for $N_c\ge3$, SE for the adjoint representation).

\vspace*{-1mm}
\section{Correlations in the bulk of the spectrum}
\vspace*{-1mm}

The full Dirac spectrum can be computed numerically using, e.g., a
special version of the Lanczos algorithm.  Which features of the
spectrum are described by RMT in the bulk of the spectrum, i.e., away
from the edges?  The global spectral density is certainly not given by
the RMT result since it is sensitive to the details of the dynamics.
However, one can separate the average spectral density from the
spectral fluctuations on the scale of the local mean level spacing.
This process is called unfolding.\cite{Guhr98} After unfolding, the
spectral correlation functions are, up to a certain energy, given by
RMT.  This limiting energy is called the Thouless energy and will be
addressed in Sec.~\ref{sec:Thouless}.

The short- and long-range correlations of the eigenvalues are measured
by quantities such as the distribution, $P(s)$, of spacings, $s$,
between adjacent eigenvalues and the number variance,
$\Sigma^2(L)=\langle (n(L)-\langle n(L)\rangle)^2\rangle$, where
$n(L)$ is the number of eigenvalues in an interval of length $L$.  In
the bulk of the spectrum, the predictions of the chiral RMT ensembles
for these quantities are identical to those of the corresponding
non-chiral ensemble.  They can be tested against lattice data.  The
predictions of RMT were confirmed with very high accuracy in the
following cases: staggered fermions in SU(2),\cite{Hala95,Guhr98} in
SU(3),\cite{Pull98} and in compact U(1),\cite{Berg98} respectively,
Wilson fermions in SU(2),\cite{Hala95} and the overlap Dirac operator
in SU(2), in SU(3), and in the adjoint representation of SU(2),
respectively.\cite{Edwa99b} Furthermore, the lattice data continue to
agree with the RMT predictions in the deconfinement
phase.\cite{Pull98} In all cases, the agreement was perfect, see
$P(s)$ in Fig.~\ref{fig1} for an example.

There is a very interesting question concerning staggered fermions:
for $N_c=2$, they have the symmetries of the chSE whereas the
continuum symmetries are those of the chOE (and the other way around
in the adjoint representation).  While one should eventually see a
transition from chSE to chOE behavior as the lattice spacing goes to
zero, it is unlikely that this can be observed on present day
lattices.  The overlap Dirac operator, on the other hand, has the
correct symmetries in all cases.

If a chemical potential is added to the problem, the Dirac eigenvalues
are scattered in the complex plane.  In this case, $P(s)$ can also be
constructed, see the talk by Markum.

\vspace*{-0.5mm}
\section{Correlations at the spectrum edge}
\vspace*{-0.5mm}

Because the nonzero eigenvalues of $i\slash{D}$ come in pairs
$\pm\lambda_n$, the Dirac spectrum has a ``hard edge'' at $\lambda=0$.
As mentioned in the introduction, the Dirac eigenvalues in this region
are of great interest because of their connection to chiral symmetry
breaking.  The distribution of the smallest eigenvalues is encoded in
the microscopic spectral density,
$\rho_s(z)=\lim_{V\to\infty}\rho(z/V\Sigma)/V\Sigma$, which is
universal and can be computed in RMT.  One can also construct the
distribution of the smallest eigenvalue, $P(\lambda_{\rm min})$, and
higher-order spectral correlation functions.  Analytical RMT results
for these quantities are available.  To compare these to lattice data,
the only input one needs is the energy scale $V\Sigma=\pi\rho(0)$,
which is obtained from the data by extracting $\rho(0)$.

The spectral correlations in the vicinity of $\lambda=0$, in contrast
to those in the bulk, are sensitive to the number of massless (or very
light) quarks and to the topological charge $\nu$.  This is another
reason why the hard edge of the spectrum is more interesting than the
bulk.

\vspace*{-0.5mm}
\subsection{Quenched approximation}
\vspace*{-0.5mm}

The first evidence from the lattice for the universality of the
microscopic spectral density came from an analysis by Verbaarschot of
Columbia data for the valence quark mass dependence of the chiral
condensate in SU(3) with staggered fermions.\cite{Verb96} Plotting all
the data computed for various values of $\beta=6/g^2$ in a rescaled
form suggested by RMT, it was observed that in the region of very
small masses they all fell on the same universal curve given by RMT.
The simulations were done with dynamical quarks, but the sea quarks
were much heavier than the valence quark so that one effectively had
$N_f=0$.

\begin{figure}[t]
  \centerline{\epsfig{figure=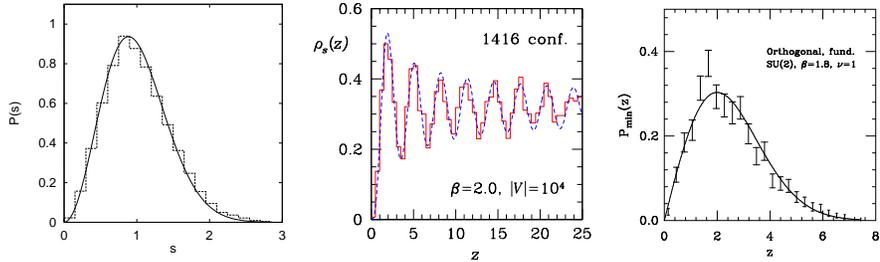,width=27.5pc}}
  \vspace*{-1mm}
  \caption{Spectral properties of the QCD lattice Dirac operator.  The
    histograms represent lattice data, the curves are RMT predictions.
    Left: Nearest neighbor spacing distribution for SU(3), staggered
    fermions (chUE), $V=6^3\times4$, $\beta=5.0$.\protect\cite{Pull98}
    Middle: Microscopic spectral density for SU(2), staggered fermions
    (chSE), $V=10^4$, $\beta=2.0$.\protect\cite{Berb98a} Right:
    Distribution of the smallest eigenvalue for SU(2), overlap Dirac
    operator (chOE), $V=4^4$, $\beta=1.8$,
    $\nu=1$.\protect\cite{Edwa99b}} \vspace*{-5mm}
  \label{fig1}
\end{figure}

The first direct lattice computation of the microscopic spectral
quantities was done in quenched SU(2) with staggered
fermions,\cite{Berb98a} and the data were subsequently analyzed in
more detail.\cite{Ma98} The agreement between lattice data and RMT
predictions was remarkably good, see $\rho_s(z)$ in Fig.~\ref{fig1}
for an example, but the analysis immediately raised a number of
questions.  First, why were the data consistent with the RMT
predictions for topological charge $\nu=0$ even in weak coupling?
Second, at what energies does RMT cease to be applicable?  The answers
will be given in Secs.~\ref{sec:topology} and \ref{sec:Thouless}.

Meanwhile, several groups have performed lattice simulations of the
hard edge of the spectrum for almost all interesting cases: SU(3) with
staggered fermions in three\cite{Damg98} and four\cite{Damg99,Gock98}
dimensions, the Schwinger model using both a fixed point and the
Neuberger Dirac operator,\cite{Farc98} SU(2) and SU(3) with staggered
fermions in the adjoint representation,\cite{Edwa99a} and the overlap
Dirac operator in SU(2), in SU(3), and in the adjoint representation
of SU(2), respectively.\cite{Edwa99b} The theoretical predictions of
all three chiral ensembles (and of the UE) have thus been verified
with high accuracy.

\vspace*{-1.3mm}
\subsection{Light dynamical fermions}
\label{sec:dyna}
\vspace*{-0.7mm}

In the chiral limit, the microscopic spectral correlations predicted
by RMT depend on the number, $N_f$, of massless dynamical quarks.  If
the dynamical quarks are sufficiently heavy, the microscopic spectral
correlations are given by the quenched result.  There is an
intermediate regime for quark masses of order $1/V\Sigma$, i.e., of
the same order as the low-lying Dirac eigenvalues.  In this
``double-scaling'' regime, the RMT results depend on the quark masses.
So far, analytical results are only known for the chUE.  Lattice
simulations with very light dynamical quarks have been performed in
SU(2) with staggered fermions,\cite{Berb98c} corresponding to the chSE
for which the corresponding RMT predictions were computed numerically.
Again, very good agreement was found for $\rho_s(z)$ and
$P(\lambda_{\rm min})$.

\subsection{Topology}
\label{sec:topology}

Lattice simulations with staggered fermions show that the microscopic
spectral quantities agree with the RMT predictions in the sector of
zero topological charge, even in weak coupling.  Presumably, the
reason is that the would-be zero modes related to topology are shifted
away from zero due to discretization errors of order $a^2$, where $a$
is the lattice spacing.  Thus, it is essential to use ``better'' Dirac
operators which obey the Ginsparg-Wilson relation and for which the
Atiyah-Singer index theorem holds.  This has recently been done, in
the Schwinger model using both a fixed point and the Neuberger Dirac
operator,\cite{Farc98} and using the overlap Dirac operator in SU(2),
in SU(3), and in the adjoint representation of SU(2),
respectively.\cite{Edwa99b} (``Neuberger'' and ``overlap'' refer to
the same operator.)  The lattice data agree very well with the RMT
predictions also in sectors with $\nu\ne0$, see $P(\lambda_{\rm min})$
in Fig.~\ref{fig1}.

\vspace*{-0.5mm}
\section{Thouless energy}
\label{sec:Thouless}
\vspace*{-0.5mm}

Since RMT predictions are derived without any knowledge of the
details of the QCD dynamics, they are only valid up to a certain
energy which is called the Thouless energy, $E_c$.  To identify $E_c$
for QCD, one needs two ingredients: (1) the fact that the random
matrix model is applicable only if the kinetic terms in the chiral
Lagrangian can be neglected, which is the case for $L\ll 1/m_\pi$
(with $L$ the linear extent of the four-volume and $m_\pi$ the pion
mass), and (2) the Gell-Mann--Oakes--Renner relation, connecting
$m_\pi$ to the quark mass and the chiral condensate.  This gives rise
to the theoretical prediction\cite{Osbo98a,Jani98} $E_c\sim
f_\pi^2/\Sigma L^2$, where $f_\pi$ is the pion decay constant.  This
prediction was verified numerically in instanton liquid
simulations\cite{Osbo98a} and in lattice simulations for
SU(2)\cite{Berb98d,Guhr98} and SU(3)\cite{Gock98} with staggered
fermions.  For lattice simulations, this roughly means that the domain
of validity of the RMT approach is larger for larger lattice size and
stronger coupling.

\vspace*{-0.5mm}
\section{Summary}
\vspace*{-0.5mm}

We now have a wealth of analytical and numerical evidence in support
of the claim that the low-lying spectrum of the QCD Dirac operator is
described by universal functions which can be computed, e.g., in RMT.
This applies to the quenched approximation, to the full theory with
dynamical quarks, and to trivial and nontrivial topological sectors.
Moreover, we know the energy where the universality breaks down.
Apart from a better analytical understanding of the Dirac spectrum,
this also offers a variety of practical applications, such as better
extrapolations to the thermodynamic limit,\cite{Berb98b} the
extraction of chiral logarithms,\cite{Berb99} and perhaps the
construction of hybrid Monte Carlo algorithms which would use the
distribution of the small eigenvalues as input.

\bigskip

\noindent {\bf Acknowledgments.}
I thank the organizers for the invitation to this very stimulating
workshop, and M.E. Berbenni-Bitsch, P.H. Damgaard, M.  G\"ocke\-ler, T.
Guhr, H.  Hehl, A.D.  Jackson, J.-Z. Ma, H. Markum, S.  Meyer, S.
Nishigaki, R.  Pullirsch, P.E.L. Rakow, A.  Sch\"afer, B.  Seif,
J.J.M. Verbaarschot, H.A.  Weidenm\"uller, and T. Wilke for very
enjoyable collaborations.

\end{document}